\documentclass[aps,reprint,prl,superscriptaddress,showpacs,floats,floatfix]{revtex4-2}
\usepackage{graphicx}
\usepackage{amsmath,graphicx,dcolumn}
\usepackage[colorlinks=true, citecolor=blue, linkcolor=blue, urlcolor=blue, citebordercolor={1 1 1}]{hyperref}
\usepackage[usenames]{color}
\usepackage{datetime}
\usepackage{textcomp}
\usepackage{ulem}
\usepackage{morefloats}
\usepackage{mathptmx}
\usepackage{indentfirst}
\usepackage{bm}
\usepackage[cmyk]{xcolor}
\usepackage{siunitx}
\usepackage{ulem}

\newcommand{\blue}{\textcolor{blue}}

\begin{document}

\title{Termination-selective Kondo hybridization of topological surface states in CeCuAs$_2$}

\author{Bo Chen}
\affiliation{Institute of Super-microstructure and Ultrafast Process in Advanced Materials, School of Physics, Central South University, Changsha 410083, Hunan, China}

\author{Chen Zhang}
\affiliation{Institute of Super-microstructure and Ultrafast Process in Advanced Materials, School of Physics, Central South University, Changsha 410083, Hunan, China}

\author{Hao Liu}
\affiliation{Institute of Super-microstructure and Ultrafast Process in Advanced Materials, School of Physics, Central South University, Changsha 410083, Hunan, China}

\author{Qi-Yi Wu}
\affiliation{Institute of Super-microstructure and Ultrafast Process in Advanced Materials, School of Physics, Central South University, Changsha 410083, Hunan, China}

\author{Zhong-Tuo Fu}
\affiliation{Institute of Super-microstructure and Ultrafast Process in Advanced Materials, School of Physics, Central South University, Changsha 410083, Hunan, China}

\author{Ying Zhou}
\affiliation{Institute of Super-microstructure and Ultrafast Process in Advanced Materials, School of Physics, Central South University, Changsha 410083, Hunan, China}

\author{Chun-Hui Lyu}
\affiliation{Institute of Super-microstructure and Ultrafast Process in Advanced Materials, School of Physics, Central South University, Changsha 410083, Hunan, China}

\author{Yang Luo}
\affiliation{Institute of Super-microstructure and Ultrafast Process in Advanced Materials, School of Physics, Central South University, Changsha 410083, Hunan, China}

\author{Zi-Jie Xu}
\affiliation{Institute of Super-microstructure and Ultrafast Process in Advanced Materials, School of Physics, Central South University, Changsha 410083, Hunan, China}

\author{Hai-Long Deng}
\affiliation{Institute of Super-microstructure and Ultrafast Process in Advanced Materials, School of Physics, Central South University, Changsha 410083, Hunan, China}

\author{Zheng-Tai Liu}
\affiliation{Shanghai Synchrotron Radiation Facility, Shanghai Advanced Research Institute, Chinese Academy of Sciences, Shanghai 201204, China}

\author{Mao Ye}
\affiliation{Shanghai Synchrotron Radiation Facility, Shanghai Advanced Research Institute, Chinese Academy of Sciences, Shanghai 201204, China}

\author{Zhen-Hua Chen}
\affiliation{Shanghai Synchrotron Radiation Facility, Shanghai Advanced Research Institute, Chinese Academy of Sciences, Shanghai 201204, China}

\author{Yao-Bo Huang}
\affiliation{Shanghai Synchrotron Radiation Facility, Shanghai Advanced Research Institute, Chinese Academy of Sciences, Shanghai 201204, China}

\author{Sheng-Tao Cui}
\affiliation{National Synchrotron Radiation Laboratory, University of Science and Technology of China, Hefei 230029, Anhui, China}

\author{Zhe Sun}
\affiliation{National Synchrotron Radiation Laboratory, University of Science and Technology of China, Hefei 230029, Anhui, China}

\author{Da-Wei Shen}
\affiliation{National Synchrotron Radiation Laboratory and School of Nuclear Science and Technology, University of Science and Technology of China, Hefei 230026, Anhui, China}

\author{Jun He}
\affiliation{Institute of Super-microstructure and Ultrafast Process in Advanced Materials, School of Physics, Central South University, Changsha 410083, Hunan, China}

\author{Yu-Xia Duan}
\affiliation{Institute of Super-microstructure and Ultrafast Process in Advanced Materials, School of Physics, Central South University, Changsha 410083, Hunan, China}

\author{Jian-Qiao Meng}
\email{Corresponding author: jqmeng@csu.edu.cn}\affiliation{Institute of Super-microstructure and Ultrafast Process in Advanced Materials, School of Physics, Central South University, Changsha 410083, Hunan, China}

\date{\today}

\begin{abstract}

Strongly correlated topological states in Kondo lattices represent a frontier in condensed matter physics, where the coupling between boundary electrons and localized f states can reshape the low-energy electronic structure. Here, we combine termination-resolved angle-resolved photoemission spectroscopy (ARPES), Ce 4$d \rightarrow 4f$ resonant photoemission, and surface electronic structure calculations to investigate the Kondo lattice compound CeCuAs$_2$. Two distinct surface electronic structures are assigned to the As1 and CeAs2 terminations through their characteristic Fermi-surface geometries and termination-specific calculations. On CeAs2, a Dirac surface state associated with the calculated nontrivial bulk topology coexists with a weakly dispersive Ce 4$f$ band near $E_F$. Resonant photoemission identifies the low-energy 4$f$ character and reveals crystal-field-related fine structure in the energy range traversed by the Dirac branch. On cooling, the 4$f$ and Dirac regions both gain spectral weight, whereas a reference conduction-band region changes much less. Together, these observations provide spectroscopic evidence for termination-selective $f$–Dirac hybridization and the development of 4$f$ character in a topological surface state. Our results identify surface termination as a microscopic means of controlling the interplay between Kondo correlations and topological boundary electrons.

\end{abstract}

\maketitle


Heavy-fermion systems host a remarkably rich spectrum of emergent quantum phenomena, encompassing unconventional superconductivity \cite{FSteglich1979PRL, MBMaple1986PRL, WSchlabitz1986ZPB}, heavy-quasiparticle formation \cite{KAndres1975PRL, GRStewart1984RMP}, pressure-tuned superconducting phases \cite{HQYuan2003Science}, non-Fermi-liquid physics \cite{VASidorov2002PRL, ZFWeng2016RPP}, and, more recently, topological electronic states \cite{MDzero2010PRL, MNeupane2013NC, GLi2014Science, JJiang2013NC, NXu2014NC, NXu2014PRB, BSTan2015Science}. The competition among low-energy electronic scales renders their phase diagrams sensitive to external tuning parameters such as pressure, magnetic field, and chemical doping, and makes their electronic structures susceptible to surface and interface modifications. Bulk topology constrains boundary states on symmetry-compatible surfaces, while their dispersion, orbital composition, Dirac-point energy, and coupling to other surface degrees of freedom can depend strongly on the microscopic termination. This raises a fundamental question: how does surface termination control the coupling between boundary electrons and many-body correlations?

Central to this inquiry is the recognition that different surface terminations\cite{JKim2014PRB, SPatil2016NC, GPoelchen2020npjQM} not only generate distinct boundary electronic structures but also modify the local coordination and orbital overlap through which surface electrons couple to the Ce 4$f$ degrees of freedom. In topological Kondo insulators \cite{MDzero2010PRL, MDzero2016ARCMP}, hybridization between itinerant and localized states can redistribute spectral weight, reshape quasiparticle dispersions, and admix $f$-electron character into boundary states. Related surface renormalization effects have been investigated in heavy-fermion compounds \cite{MHoppner2013NC, KHagiwara2016NC}. Nevertheless, experimentally connecting a termination-selected Dirac-like branch to crystal-field-related 4$f$ spectral weight and its temperature evolution remains challenging \cite{MDzero2016ARCMP, MDzero2012PRB}. This brings us to the central question: can a termination-selected Dirac-like surface state exhibit spectroscopic signatures of Kondo coupling while retaining a recognizable dispersive branch?

Recently, CeCuAs$_2$ has emerged as a promising candidate in this context. Its crystal structure features an arsenic square-net layer, a motif known to generate Dirac-like electronic structures in related compounds \cite{SDatta2023APL, JPark2011PRL, KWang2012PRB, BKang2022AM}, while the Ce 4$f$ electrons introduce strong electronic correlations. This combination motivates an examination of how the temperature-dependent development of 4$f$ hybridization affects the surface electronic structure. Theoretical studies of Kondo insulators \cite{MDzero2010PRL, JWerner2013PRB} and investigations of Ce-based heavy fermions \cite{JHShim2007Science, CXu2022npjQM} highlight the importance of the hybridization and $f$-electron delocalization  for low-energy electronic states.

\begin{figure*}[htp]
\begin{center}
\includegraphics[width=1.6\columnwidth]{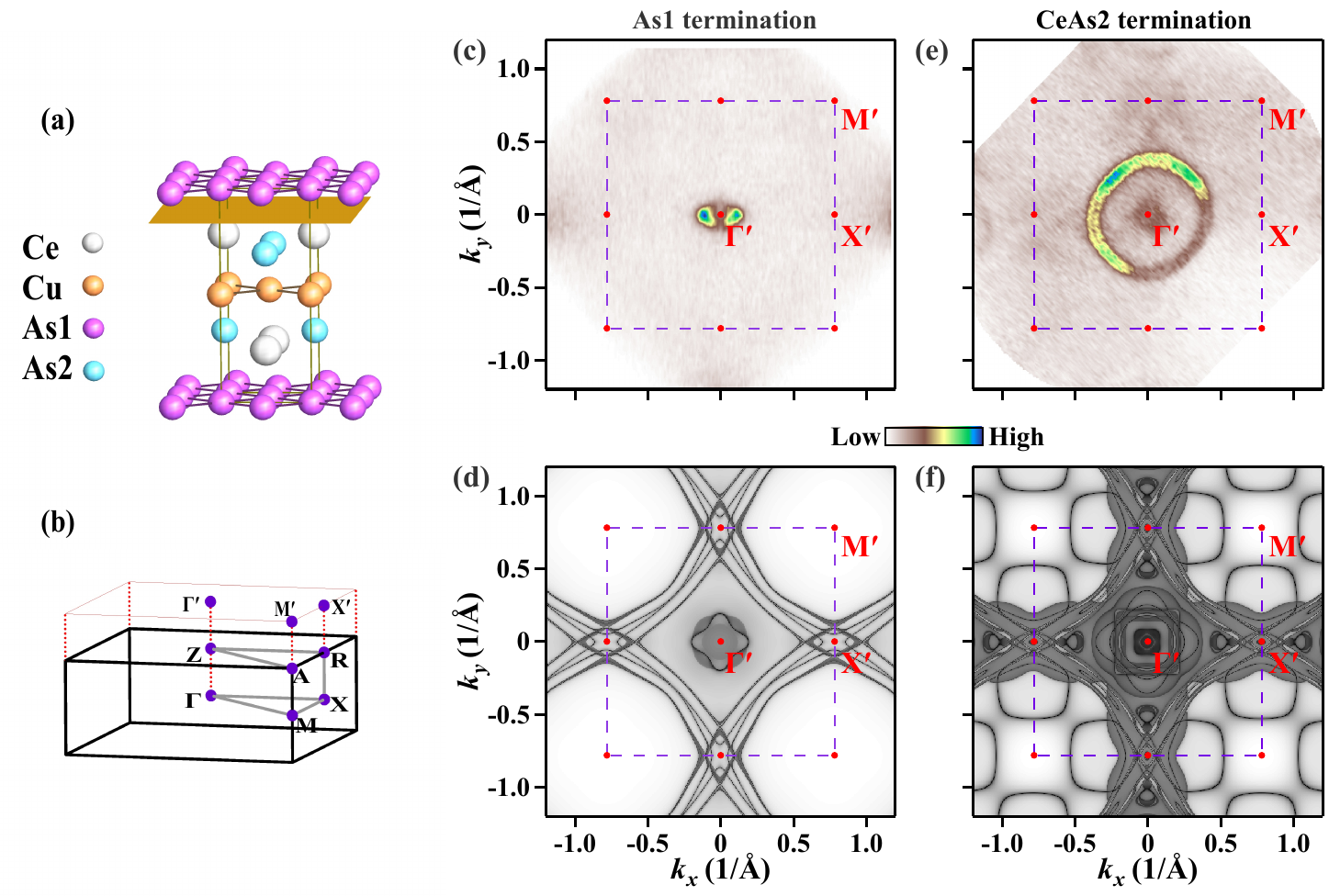}
\end{center}
\caption{(color online) Termination-resolved Fermi surfaces and surface assignment of CeCuAs$_2$. (a) Crystal structure of CeCuAs$_2$; the dark yellow plane indicates the cleavage interface. (b) Three-dimensional bulk BZ and the projected (001) surface BZ, with high-symmetry points marked. (c,e) Fermi-level $k_x$ -$k_y$ intensity maps measured at 10 K with 85 and 88 eV photons on the As1 and CeAs2 terminations, respectively, integrated within $E_F \pm$ 10 meV. (d,f) Calculated FS contours for the As1 and CeAs2 terminations, respectively, with the Ce 4$f$ electrons treated as localized core states.}\label{FIG:1}
\end{figure*}

Early studies characterized CeCuAs$_2$ as either a heavy-fermion semimetal or a Kondo semiconductor based on its semiconducting resistivity and small thermopower  \cite{KSengupta2004PRB, EVSampathkumaran2005PB, MSzlawska2008JAC, DEvans2022Thesis}. Subsequent experiments revealed large negative magnetoresistance and spin-glass-like behavior below $T_f\approx4.5$ K \cite{LChen2023IM}. Transport measurements and electronic structure calculations have motivated its identification as a topological insulator candidate, while the origin of its unusual transport remains under discussion \cite{LChen2023IM, SDatta2023APL, LCvitkovich2019Thesis}. Previous photoemission studies established surface–bulk differences in the electronic structure and hybridization \cite{SDatta2023APL, SDatta2022JPCS}. The unresolved issue is how these correlations manifest in momentum-resolved surface bands associated with different terminations.

Here, we combine termination-resolved ARPES, Ce $4d \rightarrow 4f$ resonant ARPES, and surface calculations to investigate CeCuAs$_2$. Two distinct surface electronic structures are assigned to the As1 and CeAs2 terminations through their characteristic Fermi-surface geometries and termination-specific calculations. The CeAs2 termination hosts a Dirac surface state associated with the nontrivial bulk band topology, together with a weakly dispersive 4$f$-derived band near $E_F$. The 4$f$ character is identified by resonant photoemission, and the low-energy fine structure is associated with the crystalline electric field (CEF). Their energy–momentum overlap and correlated enhancement on cooling provide evidence for $f$–Dirac hybridization. The combination of termination selectivity, orbital identification, and temperature evolution reveals how the surface environment selects a coupling channel between topological boundary electrons and the Kondo lattice.

High-quality single crystals of CeCuAs$_2$ were grown using a Bi-flux method. ARPES measurements were conducted at the BL09U and BL03U beamlines of the Shanghai Synchrotron Radiation Facility and at the BL13U beamline of the National Synchrotron Radiation Laboratory, using a Scienta DA30 analyzer under ultrahigh-vacuum conditions below $6\times10^{-11}$ mbar. All samples were cleaved $in$ $situ$ at approximately 10 K. Temperature-dependent measurements were used to examine the evolution of the surface electronic structure. For comparison with experiment, termination-specific rigid energy offsets were applied to the calculated spectra to align the reference bands with the measured electronic structure. Such offsets can arise from differences in near-surface charge redistribution and electrostatic potential between the idealized calculations and the cleaved surfaces. For each termination, the same offset was used in the Fermi-surface and band-dispersion comparisons. Details of sample characterization and electronic structure calculations are provided in the Supplemental Material \cite{SupplementalMaterial}.

\begin{figure*}[htp]
\begin{center}
\includegraphics[width=1.6\columnwidth]{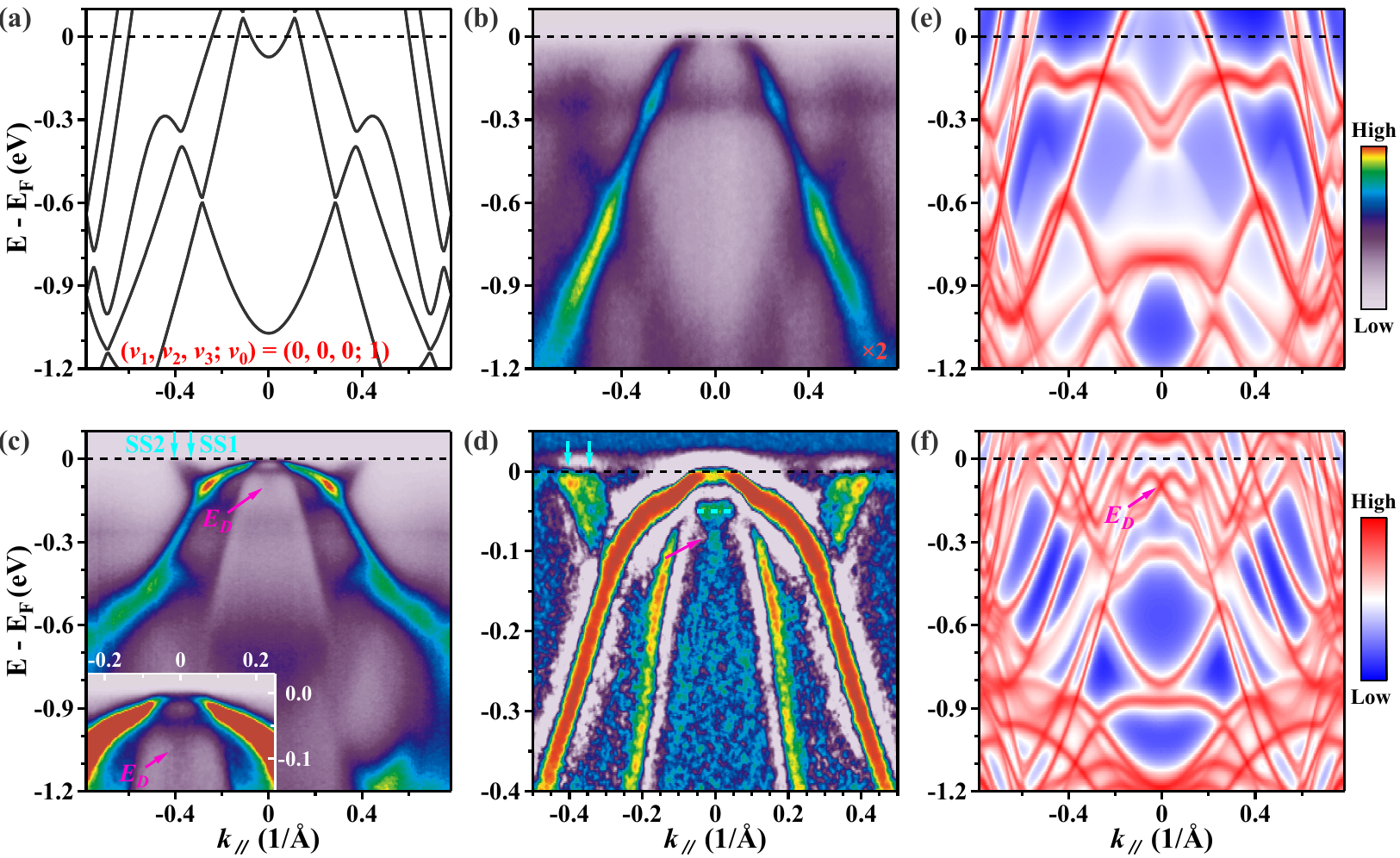}
\end{center}
\caption{Termination-dependent low-energy electronic structure of CeCuAs$_2$. (a) Calculated bulk band structure including SOC along $\Gamma$-$\rm X$. (b,c) ARPES spectra measured with 54 eV photons along $\rm X^{\prime}$-$\Gamma^{\prime}$-$\rm X^{\prime}$ on the As1 and CeAs2 terminations, respectively. SS1 and SS2 label additional features on CeAs2, and $E_D$ marks the approximate Dirac-point energy. The inset in (c) magnifies the low-energy region around $\Gamma^{\prime}$. (d) Second-derivative image of the CeAs2 spectrum in (c), highlighting the Dirac dispersion and a weakly dispersive low-energy feature around $\Gamma^{\prime}$. (e,f) Calculated surface-projected electronic structures for the As1 and CeAs2 terminations, respectively. Pink arrows indicate the Dirac-point energy $E_D$ in (c), (d), and (f).}\label{FIG:2}
\end{figure*}

CeCuAs$_2$ crystallizes in a tetragonal structure with space group $P4/nmm$ and exhibits a layered stacking sequence along the $c$ axis, as shown in Fig. \blue{1(a)}. The three-dimensional bulk Brillouin zone (BZ) and its projection onto the (001) surface BZ are shown in Fig. \blue{1(b)}. The relatively weak bonding between the Ce-containing block and the As1 square-net layer favors cleavage at this interface, exposing either an As1 or a CeAs2 termination. Previous hard x-ray photoelectron spectroscopy (HAXPES) measurements highlighted the sensitivity of the Ce-derived electronic structure to the surface environment \cite{SDatta2023APL, SDatta2022JPCS}. Distinguishing the two terminations is therefore essential for interpreting their low-energy ARPES spectra.

In our ARPES measurements at 10 K, we reproducibly observed two distinct classes of Fermi-surface (FS) patterns after cleaving different samples. Figure \blue{1(c)} displays the photoemission intensity integrated within [-10,10] meV of $E_F$, measured with 85 eV photons. The intensity is concentrated around $\Gamma^{\prime}$, with additional weaker features extending toward the surface-zone boundary. Figure \blue{1(e)}, measured with 88 eV photons, shows a markedly different pattern, including a prominent ring-like contour around $\Gamma^{\prime}$. These differences were reproducible across multiple cleaves and samples, supporting the presence of distinct surface electronic structures.

To assign these surface electronic structures, we performed density functional theory (DFT) calculations for the As1- and CeAs2-terminated (001) surfaces. The Ce 4$f$ electrons were treated as localized core states in these reference calculations. The calculated contours for As1 [Fig. \blue{1(d)}] and CeAs2 [Fig. \blue{1(f)}] capture characteristic differences between the two measured FS patterns, providing a self-consistent basis for assigning Figs. \blue{1(c)} and \blue{1(e)} to the As1 and CeAs2 terminations, respectively. These labels refer to the measured electronic structures of the corresponding terminations; individual bands need not all be purely surface localized.

We next compared the band dispersions of the two terminations and their corresponding surface calculations. CeCuAs$_2$ has been proposed as a topological insulator candidate based on its calculated SOC-induced global gap, nontrivial topological indices, and gapless surface states \cite{LChen2023IM}. Figure \blue{2(a)} shows the calculated bulk band structure including spin-orbit coupling (SOC), which provides the reference for examining the surface-projected electronic structure. The SOC-induced global gap and the strong topological indices $(\nu_1,\nu_2,\nu_3;\nu_0)=(0,0,0;1)$ identify the strong topological character of the calculated bulk bands, consistent with the previous classification of CeCuAs$_2$ \cite{LChen2023IM}. The termination-resolved surface calculations and ARPES measurements further connect this nontrivial bulk topology to the Dirac surface state observed on CeAs2. Figure \blue{2(b)} displays the ARPES intensity along $\rm X^{\prime}$-$\Gamma^{\prime}$-$\rm X^{\prime}$ for the As1 termination at 10 K. A prominent hole-like feature disperses toward $E_F$. The corresponding As1 surface-projected calculation is shown in Fig. \blue{2(e)}, using the same termination-specific energy offset as in Fig. \blue{1}. No corresponding Dirac-like crossing is resolved near $E_F$ in the measured As1 spectrum.

In contrast, the CeAs2-terminated surface exhibits markedly different low-energy electronic features. Figure \blue{2(c)} shows the ARPES dispersion along the same high-symmetry direction, where a Dirac-like branch is resolved together with additional surface states labeled SS1 and SS2. The Dirac point is located approximately 0.1 eV below $E_F$, as indicated by $E_D$. The inset of Fig. \blue{2(c)} enlarges the low-energy region around $\Gamma^{\prime}$. To highlight the dispersion near $E_D$, Fig. \blue{2(d)} presents an enlarged second-derivative image of the spectrum in Fig. \blue{2(c)}. A weakly dispersive feature, highlighted by the green dashed line, is also visible around $\Gamma^{\prime}$ at a binding energy of several tens of meV, lying just below the vertex of the primary dispersive band. As discussed below, its energy range and the resonant spectra identify a low-energy 4$f$-derived contribution. Figure \blue{2(f)} shows the corresponding surface-projected calculation for the CeAs2 termination, using the same termination-specific energy offset as in Fig. \blue{1}. The calculated Dirac point is marked by the same pink arrow used to indicate $E_D$ in the experimental spectra, facilitating direct comparison between experiment and calculation.

Photon-energy-dependent measurements provide an additional assessment of dimensionality. The As1 $k_x$-$k_z$ maps and the estimate $V_0$=15 eV are presented in Fig. \blue{S2}, while photon-energy-dependent CeAs2 dispersions and momentum distribution curves (MDCs) are shown in Fig. \blue{S3} \cite{SupplementalMaterial}. Several CeAs2-derived features retain approximately fixed in-plane momenta over the measured photon energies, consistent with weak out-of-plane dispersion. Together with their termination dependence and agreement with the surface-projected calculations, this behavior supports the surface-state assignment.

To probe the coupling between the Dirac surface state and Ce 4$f$ electrons, we employed Ce $4d \rightarrow 4f$ resonant ARPES. Figures \blue{3(a)} and \blue{3(b)} present the As1 spectra along $\rm X^{\prime}$-$\Gamma^{\prime}$-$\rm X^{\prime}$ at the off-resonance photon energy of 114 eV and the on-resonance energy of 121 eV, respectively. On As1, resonance enhances the Ce-derived emission at higher binding energy, while the near-$E_F$ fine structure remains weak. This contrasts with the pronounced low-energy response on CeAs2 and demonstrates the sensitivity of the 4$f$ spectral function to the termination environment.

\begin{figure}[tbp]
\begin{center}
\includegraphics[width=0.9\columnwidth]{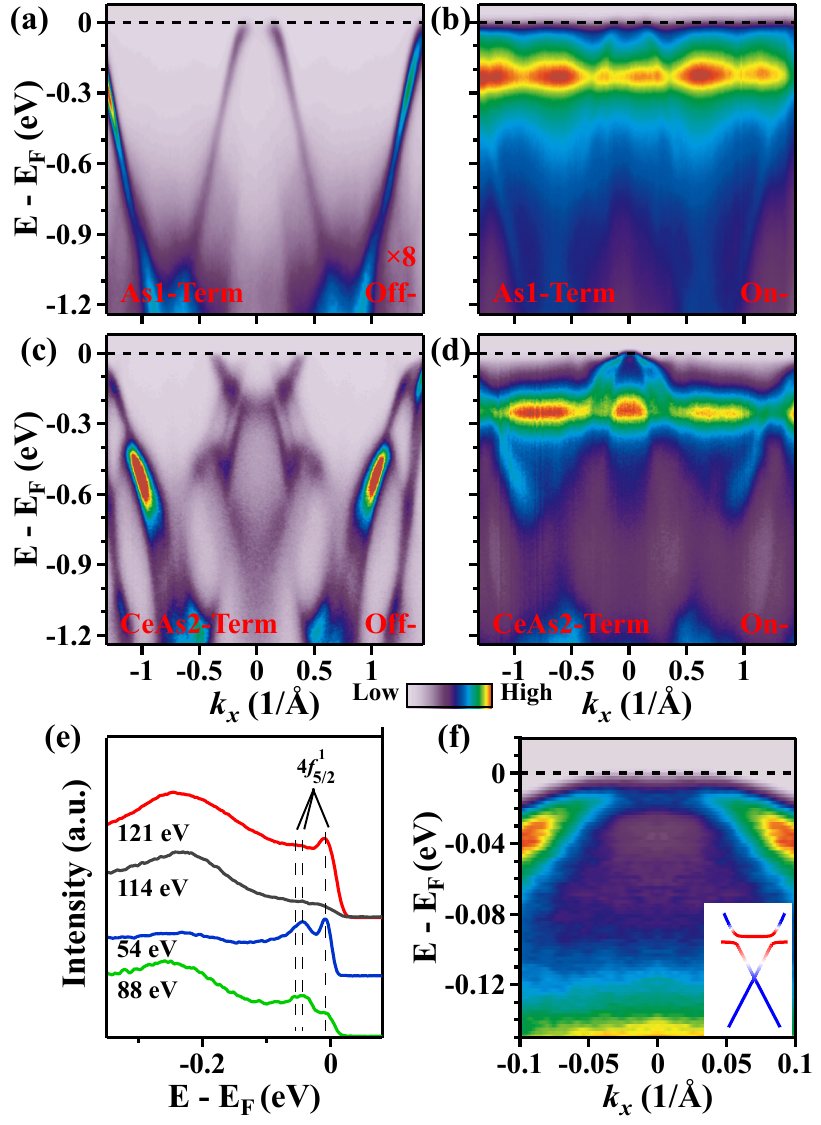}
\end{center}
\vspace*{-0.5cm}
\caption{Resonant identification of Ce 4$f$ states and their coupling to the Dirac surface band. (a,b) Off-resonance (114 eV) and on-resonance (121 eV) ARPES spectra along $\rm X^{\prime}$-$\Gamma^{\prime}$-$\rm X^{\prime}$ on the As1 termination, respectively. (c,d) Corresponding off- and on-resonance spectra on the CeAs2 termination. (e) Momentum-integrated EDCs measured with 54, 88, 114, and 121 eV photons; the near-$E_F$ 4$f^1_{5/2}$ features are indicated. (f) Enlarged on-resonance spectrum near $E_F$ and $E_D$. The inset illustrates the hybridization picture for a dispersive Dirac branch interacting with the weakly dispersive 4$f$ states.}\label{FIG:3}
\end{figure}

The CeAs2 termination exhibits a distinctly different resonant response. Figures \blue{3(c)} and \blue{3(d)} show the corresponding off- and on-resonance spectra. The near-$E_F$ 4$f$ weight is strongly enhanced at 121 eV, while the Dirac-like dispersion remains identifiable. Figure \blue{3(e)} compares energy distribution curves (EDCs) measured at selected photon energies and identifies the near-$E_F$ fine structure as Ce 4$f$-derived through its resonant enhancement. The weakly dispersive feature highlighted in Fig. \blue{2(d)} falls within the same low-energy window as the CEF-related structure resolved in the EDCs [Fig. \blue{S4}]. A near-$E_F$ flat feature is also visible in the off-resonance $\Gamma^{\prime}$- $\rm M^{\prime}$ spectrum at 54 eV and its magnified inset [Fig. \blue{S3(a)}] \cite{SupplementalMaterial}. The combined momentum-resolved and resonant evidence identifies a 4$f$-derived quasiparticle band near $E_F$, consistent with the characteristic low-energy spectra of heavy-fermion systems \cite{SPatil2016NC, SDanzenbacher2011PRL}.

\begin{figure}[tbp]
\begin{center}
\includegraphics[width=0.9\columnwidth]{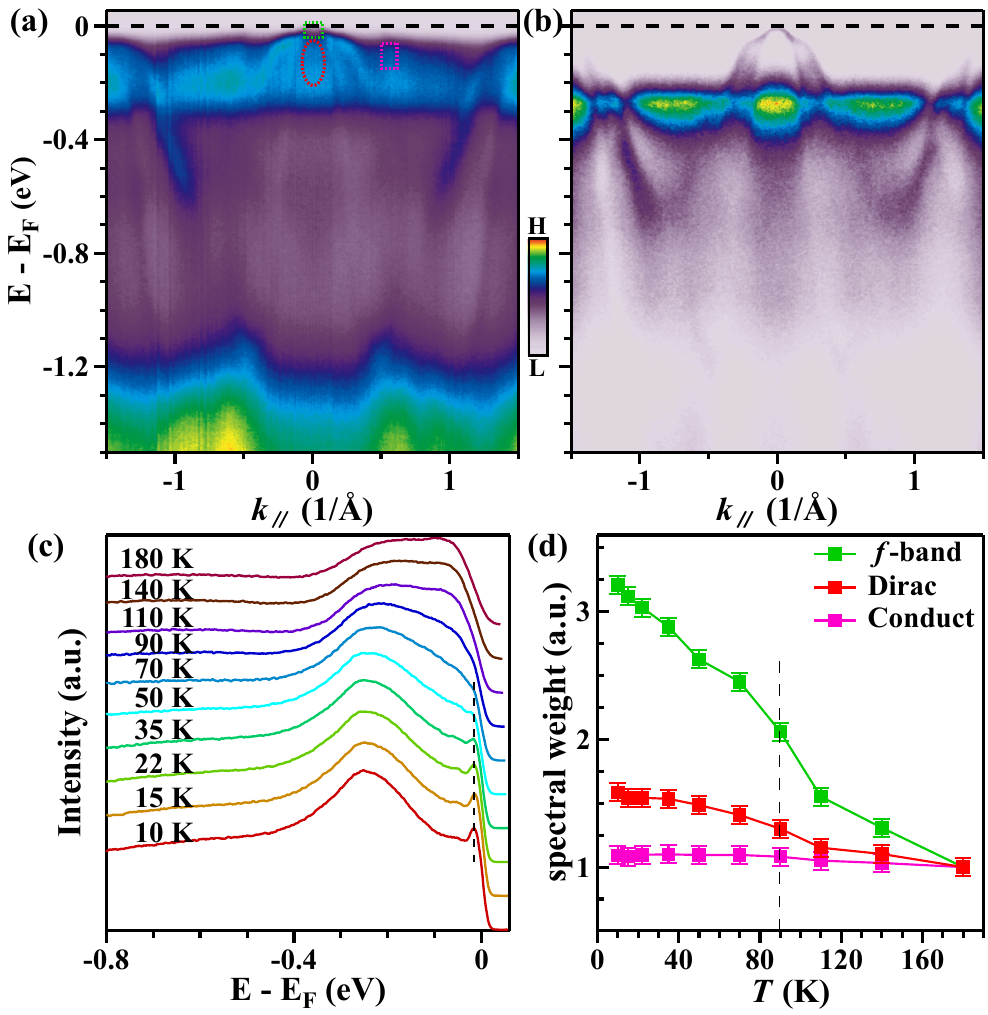}
\end{center}
\vspace*{-0.5cm}
\caption{Temperature-dependent spectral-weight evolution on the CeAs2 termination. (a) On-resonance ARPES spectrum at 110 K along $\rm X^{\prime}$-$\Gamma^{\prime}$-$\rm X^{\prime}$. The red, green, and pink regions identify the Dirac, near-$E_F$ 4$f$, and reference conduction-band ROIs, respectively. (b) On-resonance ARPES intensity difference between 10 and 110 K, $\Delta I$ = $I$(10 K) - $I$(110 K). (c) Temperature-dependent EDCs integrated within $k_\parallel\in[-0.08,0.08]$ Å$^{-1}$ around $\Gamma'$, vertically offset for clarity. (d) Integrated intensities of the three ROIs, each normalized to its corresponding value at 180 K. The dashed line marks the approximate 90 K spectroscopic  crossover scale discussed in the text.}\label{FIG:4}
\end{figure}

Figure \blue{3(f)} connects this 4$f$ assignment to the momentum structure of the Dirac surface state. The low-energy 4$f$ weight is strongly modulated in momentum in the region traversed by the dispersive branch. In conjunction with the weakly dispersive 4$f$ feature in Fig. \blue{2(d)}, this spectral structure supports hybridization between the topological surface band and the CEF-structured 4$f$ manifold. The minimal hybridization model in the Supplemental Material \cite{SupplementalMaterial} captures this orbital mixing: the Dirac-derived branch acquires 4$f$ character, accompanied by momentum-dependent spectral-weight redistribution between the hybridized branches \cite{SPatil2016NC, MHoppner2013NC}. The inset summarizes this hybridization picture.

The temperature evolution provides a further test of the many-body origin of this coupling. We therefore compare the near-$E_F$ 4$f$ region, the Dirac region, and a reference conduction-band region under the same measurement conditions.

To examine this temperature evolution, we performed systematic on-resonance ARPES measurements on the CeAs2 termination. Figure \blue{4(a)} presents the spectrum at 110 K along $\rm X^{\prime}$-$\Gamma^{\prime}$-$\rm X^{\prime}$. Three regions of interest (ROIs) mark the Dirac region (red), the near-$E_F$ 4$f$ region (green), and a reference conduction-band region (pink). Compared with the 110 K spectrum, the 10 K data in Fig. \blue{3(d)} show enhanced near-$E_F$ weight. Figure \blue{4(b)} displays the intensity difference, $\Delta I=I(10\,\mathrm{K})-I(110\,\mathrm{K})$, using the unnormalized intensity maps at both temperatures. The temperature response varies across the spectrum, with stronger changes in the low-energy regions than in the reference region.

Figure \blue{4(c)} shows temperature-dependent EDCs integrated over $k_\parallel\in[-0.08,0.08]$ Å$^{-1}$ around $\Gamma^{\prime}$. The near-$E_F$ intensity increases on cooling. The change in slope of the 4$f$-region intensity around 90 K identifies a characteristic crossover in the development of low-energy 4$f$ spectral weight \cite{QYChen2017PRB}. Figure \blue{4(d)} compares the normalized integrated intensities of the three ROIs, with each intensity normalized to its corresponding value at 180 K. The 4$f$ region exhibits the strongest relative enhancement, while the Dirac region shows a smaller but systematic increase and the reference conduction-band region remains nearly unchanged. This hierarchy is consistent with the buildup of Kondo-derived 4$f$ quasiparticle weight and its admixture into the Dirac surface band \cite{QYChen2017PRB, QYChen2018PRL,YLuo2020PRB, FYWu2023FrontPhys, YWu2021PRL, JJSong2024PRB, YZZhao2024SCPMA, BChen2024PRB}.

The joint development of the 4$f$ and Dirac-region intensities links the temperature response of the boundary band to the Kondo lattice. In a hybridized state, orbital character is shared between the dispersive and weakly dispersive branches, so the Dirac-region signal naturally contains a 4$f$ contribution. Its cooling-induced enhancement is therefore a spectroscopic signature of the coupled electronic states. Combined with the flat-band structure and resonant orbital identification in Figs. \blue{2} and \blue{3}, the contrasting ROI trends provide evidence for termination-selective Kondo hybridization of the topological surface state.

The localized-4$f$ calculations establish the termination-dependent parent surface bands and their nontrivial bulk topology, while resonant and temperature-dependent ARPES reveal how the low-energy 4$f$ degrees of freedom dress these states. The microscopic implication is that surface termination changes both the available boundary bands and their local orbital overlap with Ce 4$f$ states, thereby selecting the channel for Kondo hybridization. The persistence of a recognizable Dirac dispersion alongside the developing 4$f$ weight shows that the topological boundary band can acquire a correlated spectral character as Kondo correlations develop.

In conclusion, we have investigated the termination-dependent electronic structure of CeCuAs$_2$, a candidate correlated topological insulator. The As1 and CeAs2 surfaces are distinguished through their characteristic FS patterns and termination-specific calculations. On CeAs2, a topological Dirac surface state coexists and couples with a weakly dispersive 4$f$-derived quasiparticle band near $E_F$. Resonant orbital identification, CEF-related fine structure, and the joint enhancement of the 4$f$ and Dirac-region intensities on cooling provide spectroscopic evidence for $f$–Dirac hybridization. These results reveal surface termination as a microscopic means of selecting the coupling between topological boundary electrons and a Kondo lattice, opening a route to engineer correlated surface quasiparticles in Ce-based materials.

This work was supported by the National Natural Science Foundation of China (Grant Nos. 12574168 and 12604111), and the National Key Research and Development Program of China (Grant No. 2022YFA1604204). We are grateful for resources from the High Performance Computing Center of Central South University.

\end{document}